# Urban Magnetism Through The Lens of Geo-tagged Photography

Silvia Paldino[1], Iva Bojic[2], Stanislav Sobolevsky[21], Carlo Ratti[2], Marta C. González[3]

[1]Department of Physics,
University of Calabria,
Via Pietro Bucci, 87036 Arcavacata, Rende CS, Italy

[2]SENSEable City Laboratory,
Massachusetts Institute of Technology,
77 Massachusetts Avenue, Cambridge, 02139, USA

[3]Department of Civil and Environmental Engineering,
Massachusetts Institute of Technology,
77 Massachusetts Avenue, Cambridge, MA 02139, USA

**Abstract**

There is an increasing trend of people leaving digital traces through social media. This reality opens new horizons for urban studies. With this kind of data, researchers and urban planners can detect many aspects of how people live in cities and can also suggest how to transform cities into more efficient and smarter places to live in. In particular, their digital trails can be used to investigate tastes of individuals, and what attracts them to live in a particular city or to spend their vacation there. In this paper we propose an unconventional way to study how people experience the city, using information from geotagged photographs that people take at different locations. We compare the spatial behavior of residents and tourists in 10 most photographed cities all around the world. The study was conducted on both a global and local level. On the global scale we analyze 10 most photographed cities and measure how attractive each city is for people visiting it from other cities within the same country or from abroad. For the purpose of our analysis we construct the users' mobility network and measure the strength of the links between each pair of cities as a level of attraction of people living in one city (i.e., origin) to the other city (i.e., destination). On the local level we study the spatial distribution of user activity and identify the photographed hotspots inside each city. The proposed methodology and the results of our study are a low cost mean to characterize touristic activity within a certain location and can help cities strengthening their touristic potential.

## 1 Introduction

The traces of communication and information technologies are currently considered to be an efficient and consolidated way of collecting useful and large data sources for urban studies. There are in fact various ways to electronically track human behavior and the most diffuse one is collecting data from mobile phones [1], [2]. It was already demonstrated that this technique can be used as an accurate method for understanding crowds [1] and individual mobility patterns [3], [4], [5], to classify how the land is used [6], [7], [8], [9] or to delineate the regional boundaries [10], [11], [12]. Moreover, it was shown how to identify some of user characteristics from mobile call patterns, for example how to determinate if a user is a tourist or a resident [13]. However, when it comes to studying human mobility patterns, exploring detailed call records is not the only possibility – other sources of big data collected from digital maps [14], electronic toll systems [15], credit cards payments [16], [17], Twitter [18], circulation of bank notes [19], vehicle GPS traces [20] and also using geotagged photographs [21], [22], [23] can be successfully applied.

---

[1] Correspondence: stanly@mit.edu

The focus of this paper is on geotagged photographs that provide novel insights into how people visit and experience a city, revealing aspects of mobility and tourism, and discovering the attractions in the urban landscape. In the past, photography was already considered as a good mean of inquiry in architecture and urban planning, being used for understanding landscapes [24]. Moreover, Girardin et al. showed that it was possible to define a measure of city attractiveness by exploring big data from photograph sharing websites [25]. They analyzed two types of digital footprints generated by mobile phones that were in physical proximity to the New York City Waterfalls: cellular network activity from AT&T and photographic activity from Flickr. They distinguished between attractiveness and popularity. Regarding attractiveness they defined the Comparative Relative Strength indicator to compare the activity in one area of interest with respect to the overall activity of the city. They measured the attractiveness of a particular event in New York City. In this study we consider the attractiveness in the overall area of the city during three years, comparing the attractiveness and spatial distribution of activities in different cities.

Discovering how to increase the global city attractiveness or the local attractiveness of hotspots requires knowing the differences in the visits made by residents and tourists. While both residents and tourists take photographs at locations that they consider important, the reasons why they are taking photographs are different. This knowledge helps us to understand the different usages of the urban infrastructure in people spare time. The overall goal is to find the ways how passively collected data can be used for low cost applications that inform urban innovation. These information trends can be of interest for planning, forecasting of economic activity, tourism, or transportation [26]. Finally, a comparative study of cities from different parts of the world is a relevant objective to discern how the patterns of human behavior largely depend on a particular city [27].

In this paper we define city global attractiveness as the absolute number of photographs taken in a city by tourists, while local attractiveness of hotspots within a city is defined by the spatial distribution of photographs taken by all users (either local residents or tourists). In our analysis we are using a dataset that consist of more than 100 million publicly shared geotagged photographs took during a period of 10 years. The dataset is divided into 8,910 files denoting 3,015 different locations (e.g., cities or certain areas of interest such as Niagara Falls) where for almost every location three different labels are given: *resident*, *tourist* and *unknown,* to denote people who are living in the area, visiting the area or are uncategorized.

**2 Data**

For each photograph in the dataset, which is publicly available on the website sfgeo.org, the following data is given: user id, timestamp, geo coordinates and link from which the photograph can be downloaded. From the given dataset we omitted duplicates (9.33% of the dataset in total) together with the photographs with incorrect timestamps (0.01% of the dataset in total). In the end we were left with more than 90 million photographs. Figure 1 shows how the number of taken photographs and users changed over the period of 10 years. We can see that almost 75% of the photographs were taken from 2007 until 2010. In the further analysis we are thus considering only photographs taken during those three years (about 70 million of photographs in total).

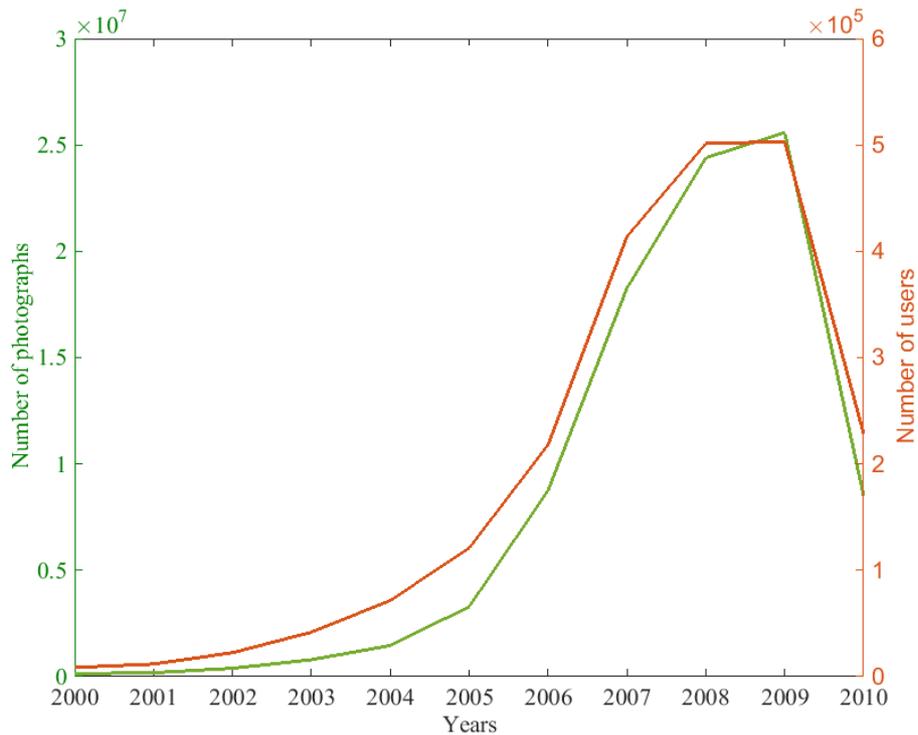

Figure 1 - Number of geotagged photographs (green line) and users (red line) from 2000 until 2010.

**3 Definition of user home cities and countries**

In order to determinate if there is any difference between how residents vs. tourists are attracted to a certain location, for each user in the dataset we have to determine his/her home city and country. Even though the dataset has tags for: *resident*, *tourist* and *unknown*, the given categorization is not comprehensive and in some cases is not consistent. For instance, for more than 85% of users in the dataset their home city is not defined (i.e., their photographs are always in *unknown* files). In addition, for almost 25% of users for who their home city is defined at least two or more cities are listed as their home cities making the proposed categorization inconsistent.

Due to the aforementioned reasons, we used our own criteria to determinate if a person is living in the area where he/she took a photograph or not. We are considering that a person is a resident if at that location he/she took the highest number of the photographs (at least 10 of them) over the longest period of time (at least longer than 180 days) calculated as the time between the first and last photograph taken at the location. Once when we determine a user home city, he/she is automatically becoming a tourist in all other cities in the dataset. A category "tourist" in this sense denotes many different kinds of visitors including business visitors. However, most of people taking photographs at locations other than their home cities in fact act like tourists during this particular instance of time.

From almost 1 million users that took photographs between 2007 and 2010, for only 11% of them we were able to determinate their home city and country using our criteria. However, these users took more than 40% of all photographs (i.e., almost 30 million). Our classification was not consistent with the initial categorization (i.e., users whose photographs were listed in only one *resident* file) for only less than 2% of users. Moreover, for every city in the dataset we identified its country code allowing us to classify tourists as domestic or foreign ones, where domestic tourists are coming from the same country as the considered city, while foreign tourists are all the others – visitors from different countries. Finally, for every city all the observed activity is assigned to residents, domestic tourists or foreign tourists and for each of these categories, we keep the following data: user id, geo coordinates and for tourists their home city id together with their country code and continent id.

# 4 Attractiveness

## 4.1 Global Attractiveness

Considering our original question what attracts people to a certain location, we start our analysis looking at different locations and their absolute global attractiveness that is quantified by the number of photographs taken in them by either domestic or foreign tourists, leaving out the contribution that was made by their residents. Once we determined user home cities and countries, in order to calculate global attractiveness for different locations, we ranked locations by the total number of photographs taken in them by tourists (i.e., people residing outside the considered city) from all over the world. We find that the first 10 ranked cities by photographs are: New York City, London, Paris, San Francisco, Washington, Barcelona, Chicago, Los Angeles, Rome and Berlin. In order to see how strong might be the impact of short-distance domestic visitors on this ranking on this classification, we compared it with the ranking built based just on the activity of foreign users in a city. Surprisingly the difference is not that high – New York and London are still the two leading cities (just switching order), Paris and San Francisco are still within top 5, while Rome having the lowest place in this new ranking among all the cities mentioned is still the 23rd world most photographed city with respect to the activity of foreign tourists. That is why the cities we picked up are the important destinations not only for all (including domestic), but also for the foreign visitors.

This ranking is also highly consistent with the one presented in [22] – all of our top 10 cities happen to be among the first 15 cities they mention. Another two global rankings of city visitor attractiveness worth mentioning in that context are the ones presented by Euromonitor[3] and MasterCard[4]. Although one should not really expect them to be consistent with our ranking, as those rankings are built on diverse (and sometimes heterogeneous) sources of data trying to include all the visits and not necessary only tourists who are willing to take photographs as we do it in our study, we will compare them against our ranking. One could often expect one city to attract more people, but another one, attracting less, being more picturesque, and motivating those fewer people attracted for taking more photos, which would result in a higher total photographic activity. However, all 10 of our top cities are included in Euromonitor's top destinations list. Worth mentioning is that this is already not the case for a newer version Euromonitor's ranking from 2015[5] – for example Washington falls out of the top 100 world destinations according to their recent estimate. This serves as a good example of how dynamic the world is, while it is not too surprising that our ranking built based on the data before 2010 happens to be more consistent with the older version of Euromonitor's report. Morover we found that our top 3 cities – New York, London and Paris - are also the top 3 (the order is different however) in MasterCard's ranking, while in total 7 of out top 10 cities (besides Berlin, Washington and Chicago) are mentioned among the "Global Top 20 Destination Cities by International Overnight Visitor Spend" in 2014.

Further, we focus this study on the global attractiveness of these 5 US cities and 5 European Union cities (EU), and add together the remaining information as the rest of Europe, the rest of the US and the rest of world. From the aforementioned ranking we were able to extract the origin-destination (O/D) network between 10 most photographed cities, the rest of the US, and Europe as well as the rest of the world. Figure 2 shows O/D flows among the top 10 city-to-city estimated flows of visitors. Colors of the ribbons correspond to destination cities and origin cities are marked with a thin stripe at the end of a ribbon (visualization method based on Krzywinski [28]). With this visualization it is evident that the most attractive cities are London and New York City, followed by San Francisco, Washington and Paris. It is interesting to point out that the most important flows happen exactly between these cities and New York City (New York City – Washington, New York City – San Francisco, New York City – London, New York City - Paris), while London and Paris have the strongest flow between each other. The less active cities are Berlin and Rome.

---

[3] http://blog.euromonitor.com/2010/01/euromonitor-internationals-top-city-destination-ranking.html
[4] http://newsroom.mastercard.com/wp-content/uploads/2014/07/Mastercard_GDCI_2014_Letter_Final_70814.pdf
[5] http://blog.euromonitor.com/2015/01/top-100-city-destinations-ranking.html

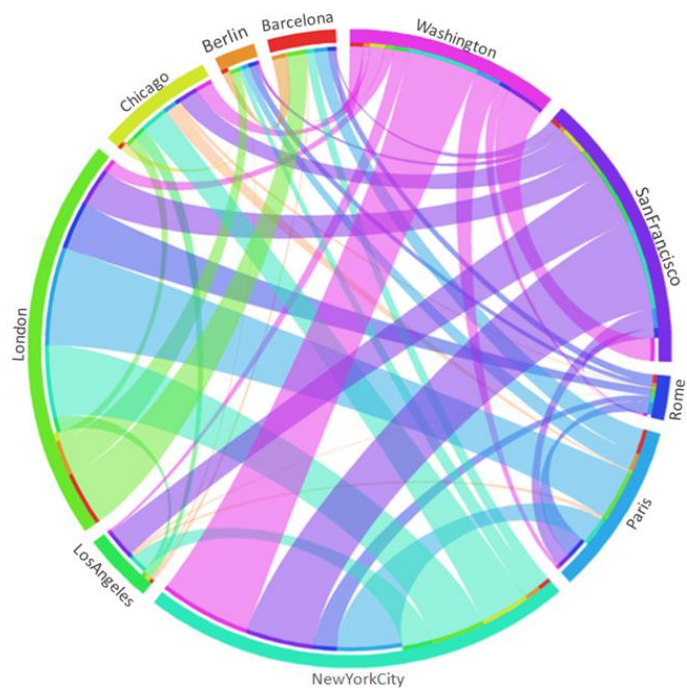

Figure 2 – Origin-destination (O/D) network among the top 10 city-to-city estimated flows of visitors. **Ribbons represent "outgoing" fluxes on each side. For example, following the ribbon between London - New York City, on London side it represents people living in London who visit and take photographs in New York City.**

The dependence between the number of users and the number of photographs they took is quite close to linear, having $R^2 \cong 0.87$ for the linear approximation. Therefore, the analysis based on the number of taken photographs is to some extent also a good proxy for the number of active users; further we will use the number of photographs as the main measure of attractiveness. However, using this measure one should also be aware of the heterogeneity of Flickr activity belonging to users coming from different parts of the world. As shown in Table 1 this heterogeneity is quite noticeable – number of photographs taken worldwide by users originating from different cities across the globe varies from 1.16 for the rest of the world to more than 1000 for San Francisco. In many places across the world people are almost never using geotagging location services (e.g., Flickr) and are not sharing their photographs publicly because of different reasons: e.g., technical, cultural or even political ones. Among the top 10 ranked cities this activity varies in a magnitude of 20 times between people living in San Francisco (the most active population) and Rome (the least active).

Table 1 – **Heterogeneity of Flickr usage: total number of photographs taken worldwide by residents of different areas versus their official population in 2008.**

| City | Population (mln) | Photographs taken | Photographs per 1000 residents |
|---|---|---|---|
| **New York City** | 8.36 | 1,026,199 | 122.75 |
| **London** | 7.81 | 1,151,799 | 147.48 |
| **Paris** | 2.23 | 534,092 | 239.50 |
| **San Francisco** | 0.81 | 851,425 | 1,051.14 |
| **Washington** | 0.59 | 525,313 | 890.36 |
| **Barcelona** | 1.62 | 255,038 | 157.43 |
| **Chicago** | 2.85 | 412,246 | 144.65 |
| **Los Angeles** | 3.83 | 289,810 | 75.67 |
| **Rome** | 2.71 | 126,011 | 46.50 |
| **Berlin** | 3.43 | 182,325 | 53.16 |
| **Rest of EU** | 4,82.61 | 8,637,148 | 17.90 |
| **Rest of the US** | 2,87.61 | 7,347,003 | 25.55 |
| **Rest of the world** | 5,905.14 | 6,877,894 | 1.16 |

In order to use the O/D flows of geotagged photography as a proxy for actual human mobility between cities across the globe, the appropriate normalization for the above heterogeneity is required. One way of doing it is by normalizing the O/D flows shown in Figure 2 by the number of photographs taken per 1000 of residents of the origin location reported in Table 1. However, this would require further assumptions about the homogeneity of the dataset representativeness for different modes of people travel behavior that would be a questionable assumption given the dataset sparsity and heterogeneity. Therefore, in the further analysis we will refrain from extrapolating the original values of O/D flows defined by the actual number of photographs taken to represent actual human mobility. In this way we will focus our analysis on the actual photographic activity of the users, keeping in mind that flows of the activity from different origins might actually have different representativeness across the entire human population and might not represent the entire variety of types of human activity from the considered origins in the considered destinations. However, we believe that photographic activity by itself is an important component of visitor behavior and might serve as a relevant proxy for measuring city visual attractiveness for the visitors.

The cumulative incoming flow for each destination in the O/D network from all the origins other than the considered destination represents the destination's total global attractiveness in terms of geotagged photographic activity of tourists. Normalized by the population of destination, this measure becomes a relative global attractiveness of the destination stating how much visitors per capita of residential population the location has.

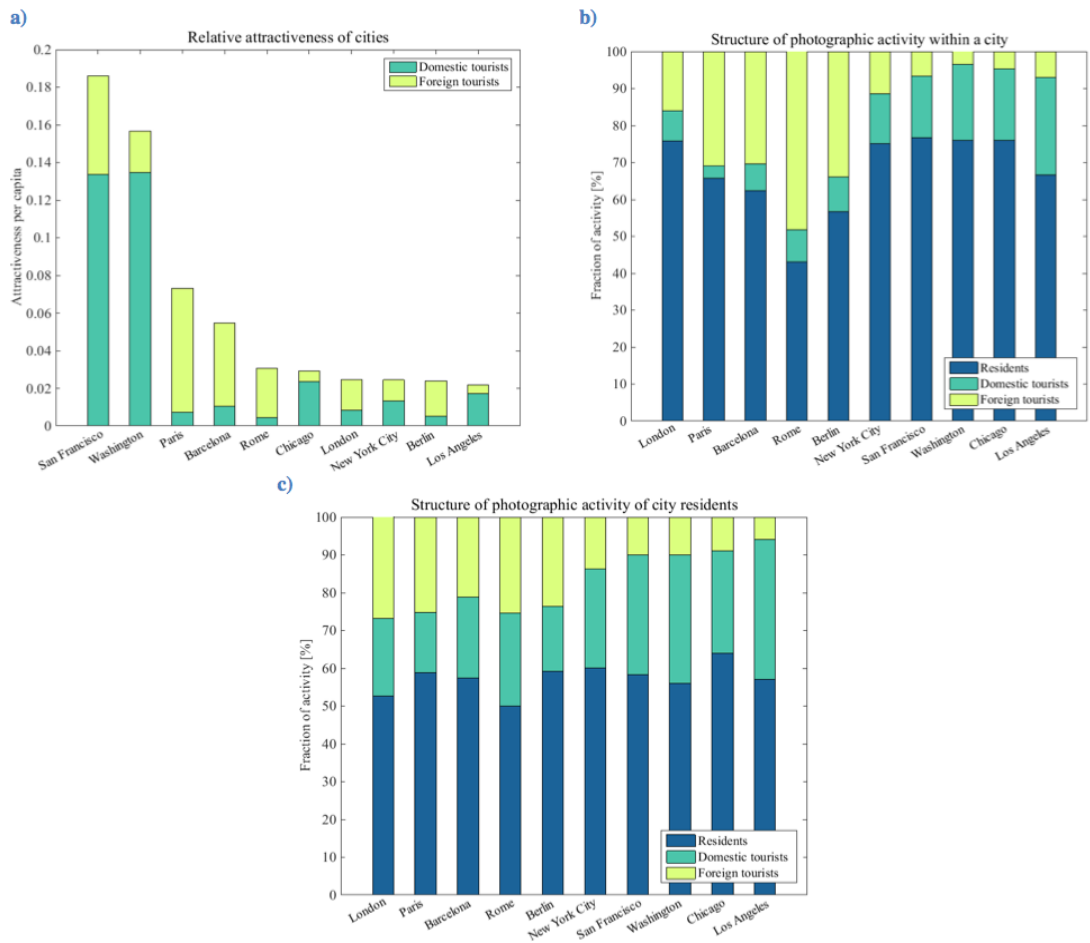

**Figure 3 – Relative attractiveness.** (a) A relative global attractiveness of top 10 cities measured as the estimated amount of touristic activity per capita of urban population, it is plotted in the decreasing order of total attractiveness, (b) the structure of photographic activity within a city and (c) the structure of photographic activity of city residents. Figure 3 (a) shows the number of domestic and foreign photographs divided by the city population (i.e., attractiveness per capita for domestic and foreign tourists). Figure 3 (b) shows the percentage of photographs within each city depending on the type of user, while Figure 3 (c) shows the share of activity type of users that reside in the particular city. For example, as depicted in Figure 3 (b), Rome receives the largest amount of foreign tourist activity and Washington the smallest. And as depicted in Figure 3 (c) people living in Washington have the largest share of domestic tourists while Parisians have the smallest.

Figure 3a shows the relative global attractiveness in relation to the population of the city (with additional distinction by domestic and foreign tourists). This gives a completely different city ranking from the initial one (e.g., New York City, London, Paris, San Francisco, Washington, Barcelona, Chicago, Los Angeles, Rome and Berlin) – now the top relative attractiveness is observed in San Francisco and in Washington for domestic tourists, while the top destination by relative attractiveness for the foreign tourists is observed in Paris. Zooming into the relative structure of photographic activity within each city (Figure 3b), a clearly distinctive pattern between European and American cities can be observed. Namely, in American cities the city residents take most of the photographs; followed by the domestic tourists mostly taking the rest, while in EU cities the activity within the cities is much more diverse showing a higher fraction of touristic and specifically foreign touristic activity. The borderline cases from both groups are London and New York City, the former having the highest fraction of residential activity among European cities making it more similar to the American cities, while the latter has the highest fraction of foreign activity among American cities, making it more similar to European ones. One of the possible explanations could be that the observed pattern is attributed to the cultural similarities going back in the history of British-American ties.

Additionally, for every city we can define a measure that shows how mobile its residents are (i.e., if their activity is home-oriented or not) by looking at the ratio of loop edges to the total outgoing weights from the O/D matrix (i.e., user activities in their home cities compared to their total activities). We depict these results in Figure 3c. Although this ratio is nearly flat varying between 50-60%, an interesting pattern appears when looking at the destinations of activities. Again, American and European patterns are surprisingly distinctive – while the American tourists seem to be mostly engaged in domestic tourism, EU citizens' travel more abroad. This difference can be explained because the US is much bigger and at the same time much more geographically diverse when compared to the EU countries of our studied cities. American users thus have more options when engaged in domestic tourism and are consequently less likely to travel to foreign destinations. When considering the strength of each particular O/D flow $a(i,j)$ between the origin $i$ and destination $j$ (i.e., a number of photographs taken by the residents of $i$ when visiting city $j$), one should expect that the number of people travelling between larger or more significant cities is higher. Therefore, in order to estimate the qualitative strength of each particular O/D flow beyond just the effect of scale, we compare all the non-loop edges of the normalized network versus the homogenous null-model where all the outgoing mobility is distributed in the relative proportion to the destination attractiveness, i.e., for each non-loop edge ($i \neq j$) the null model is:

$$model(i,j) = \frac{w^{out*}(i)\, w^{in*}(j)}{\sum_{k \neq i}(w^{in*}(k))}$$

where $w^{out/in*}(i) = w^{out/in}(i) - a(i,i)$, $w^{in}(c)$ is the total number of photographs taken in city $c$ and $w^{out}(c)$ is the total number of photographs taken worldwide by city $c$ residents. Using the aforementioned model, we compute the fraction $a(i,j)/model(i,j)$ (see Table 2) in order to see how qualitatively strong the links between each pair of cities are, i.e., how people from each origin are attracted to each destination beyond just the effect of scale.

Table 2 – **Relative strength of the links between each pair of cities, normalized by the null-model estimation.**

| O / D --> | New York City | London | Paris | San Francisco | Washington | Barcelona | Chicago | Los Angeles | Rome | Berlin |
|---|---|---|---|---|---|---|---|---|---|---|
| **New York City** | - | 0.99 | 0.88 | 1.11 | 2.08 | 0.32 | 1.60 | 0.79 | 0.61 | 0.53 |
| **London** | 0.92 | - | 1.57 | 0.73 | 0.35 | 1.59 | 0.31 | 0.44 | 1.50 | 1.37 |
| **Paris** | 0.76 | 2.44 | - | 0.60 | 0.21 | 1.60 | 0.13 | 0.24 | 1.41 | 1.60 |
| **San Francisco** | 1.58 | 0.83 | 0.44 | - | 0.86 | 0.45 | 1.46 | 2.43 | 0.41 | 0.32 |
| **Washington** | 1.70 | 0.38 | 0.77 | 1.40 | - | 0.75 | 1.07 | 0.89 | 0.67 | 0.51 |
| **Barcelona** | 0.67 | 2.40 | 1.29 | 0.20 | 0.05 | - | 0.12 | 0.19 | 1.24 | 3.32 |
| **Chicago** | 1.25 | 0.77 | 0.78 | 1.37 | 1.62 | 0.58 | - | 1.37 | 0.41 | 0.34 |
| **Los Angeles** | 1.31 | 0.56 | 0.24 | 3.31 | 0.39 | 0.50 | 1.07 | - | 0.31 | 0.31 |
| **Rome** | 0.76 | 1.63 | 1.70 | 0.21 | 0.29 | 2.25 | 0.36 | 0.07 | - | 1.69 |
| **Berlin** | 0.64 | 1.79 | 1.15 | 0.61 | 0.37 | 2.15 | 0.16 | 0.09 | 2.22 | - |

One clear pattern, which can be immediately recognized, is a high level of attractiveness between pairs of American cities, as well as a lower level of connectivity between American cities and EU ones (see Figure 4a). This pattern was rather expected, as trans-Atlantic links are likely to be weaker than more local ones. To a certain extent it could be attributed to a general decay of links with distances (similar to Krings [29]) – indeed the overall trend is well approximated by an exponential function (Figure 4b) going sharply down once when crossing the ocean. However, an interesting finding shown in both Figures 4-a and 4b is that the links going from the US to EU are stronger than same distance links going in the reciprocal direction from EU to the US (see Figure 4a). According to this, Americans are more attracted by EU destinations than Europeans are by the US cities.

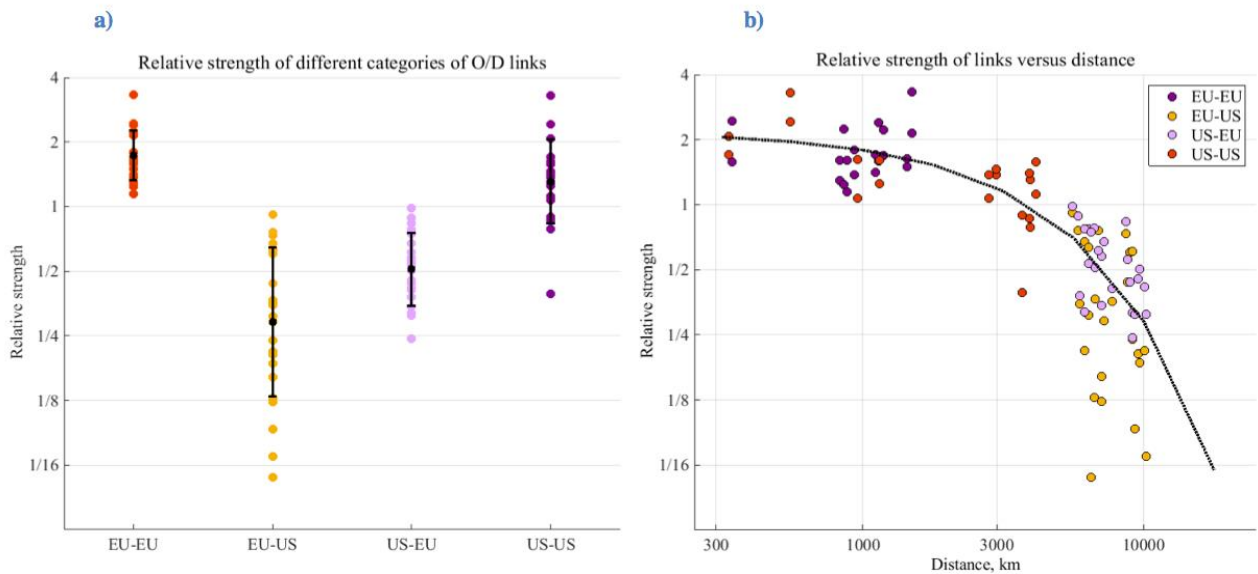

**Figure 4 –** **Relative strength of different types of links (a) grouped by continents, (b) ordered by distance.**

### 4.2 Local Attractiveness

Until this point, cities have been considered only as aggregated spatial units. However, cities all around the world are not spatially homogenous and there are always relatively more attractive areas within them to which one can refer as "hotspots". Moreover, attractiveness of different locations across the city could vary for different categories of users. In order to investigate this local attractiveness, we identified spatial distribution of photographs taken by users belonging to resident, domestic or foreign tourist groups. Visualizing them on the map, where each dot represents a photograph taken, Figure 5 shows different spatial patterns of how different categories of users take photographs in New York City, which was taken as an example. As expected, tourists in general take more photographs in central areas within the city while residents take photographs at locations that are much more spatially distributed – scattered all over the city. By putting those dots onto the map, we can create "maps of attractiveness" for every city and for its residents and tourists.

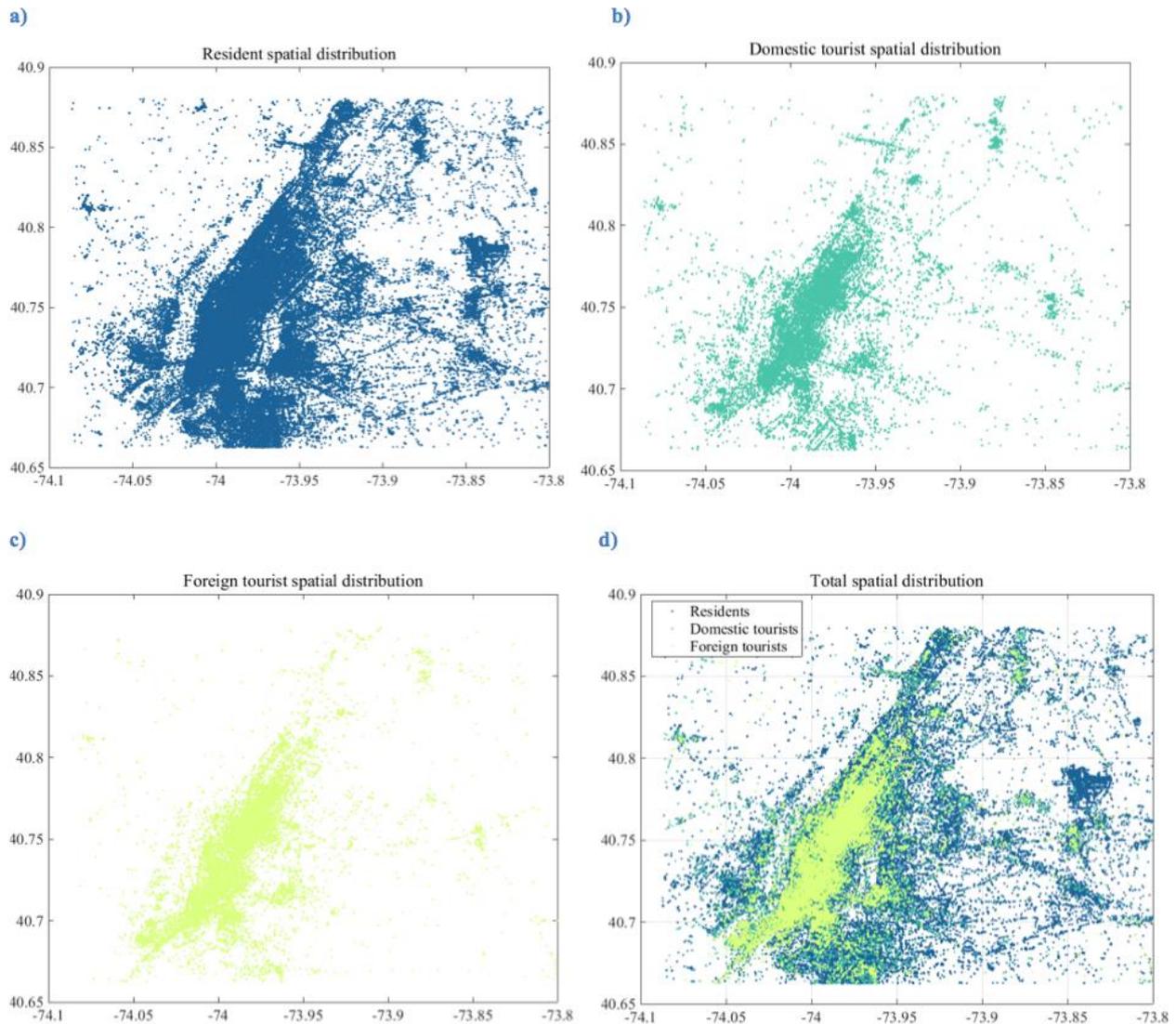

**Figure 5 – Spatial distribution in New York City.** Dots represent photographs taken by New York City (a) residents, (b) domestic tourists, (c) foreign tourists creating a map of attractiveness for three different categories, while (d) is the map of attractiveness for the three contributions together.

In order to analyze the spatial activity we use 500x500m rectangular grids covering each city. First we start with a standard analysis of the distribution functions of the activity density values across the grid cells for different categories of users – residents, domestic and foreign visitors, as well as for the total activity. All of activities can be fitted to truncated lognormal distributions (distribution truncation is an important consideration from a technical standpoint as by its definition the cells covered by user activity cannot have less than one photograph taken inside of them) as shown in Figure 6 for New York City. We took New York City just as an example as patterns for the other cities are similar with the only difference in the relative position of the curves corresponding to the foreign and domestic visitors.

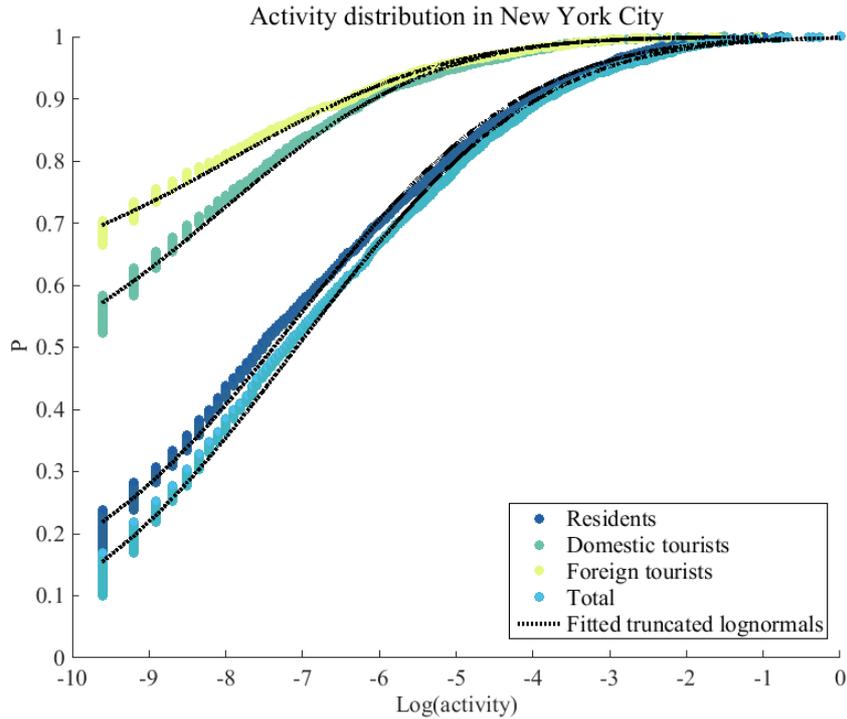

**Figure 6 – Cumulative distribution functions of activity density.** The plot shows the cumulative distribution functions of activity density over 500x500m grid for city residents, domestic and foreign visitors as well as the total activity in New York City fitted to truncated log-normal distributions.

Log-normal distribution is characterized by its mean and variance. While means vary for different types of users due to the differences of their total activity level, variance characterizes how broad or narrow the distribution is, i.e., how strongly the activity density varies over the area of the city covered by the considered user activity. The values of variance for all 10 cities and different categories of users reported in Table 3 show a consistent pattern – with the only exception of Los Angeles variance of domestic visitor where their spatial activity is always the lowest one, while variance of the foreign activity is higher for all the EU cities compared to the residential activity and lower for all the American cities. Variance of the total activity is usually the highest one with the exception of three EU cities – London, Barcelona and Rome.

**Table 3 – Variance of the fitted lognormal distributions.** Variance of the fitted lognormal distributions to the activity density is shown in 500x500m grid for city residents, domestic and foreign visitors, as well as for the total activity in the different cities.

| *City* | *Residents* | *Domestic tourists* | *Foreign tourists* | *Total activity* |
|---|---|---|---|---|
| **New York City** | 2.37 | 2.17 | 2.26 | 2.44 |
| **London** | 1.86 | 1.89 | 2.22 | 1.92 |
| **Paris** | 2.24 | 1.62 | 2.33 | 2.39 |
| **San Francisco** | 2.19 | 2.07 | 2.09 | 2.22 |
| **Washington** | 2.21 | 2.10 | 2.17 | 2.22 |
| **Barcelona** | 2.20 | 1.77 | 2.42 | 2.38 |
| **Chicago** | 2.17 | 2.03 | 2.08 | 2.23 |
| **Los Angeles** | 1.86 | 1.84 | 1.71 | 1.91 |
| **Rome** | 2.04 | 1.88 | 2.95 | 2.39 |
| **Berlin** | 1.86 | 1.78 | 2.04 | 2.07 |

Another perspective of user activity spatial distribution across different cities can be expressed by considering the dimensions of the areas covered by different quintiles of the top density cells. Figure 7 visualizes the shape of the curve characterizing those distributions (normalization is performed by the number of cells covering 50% of the total activity in order to bring different distributions on the same scale) in all 10 cities for all types of users (i.e., residents, domestic and foreign tourists). The curve shows how large is the most photographed area covering a given quintile of the total activity within the city in relation to the total area covered. Interestingly, curves for all 10 cities and three different types of users nearly follow one single shape which seems to represent a universal pattern of spatial distribution of the photographic activity across the city.

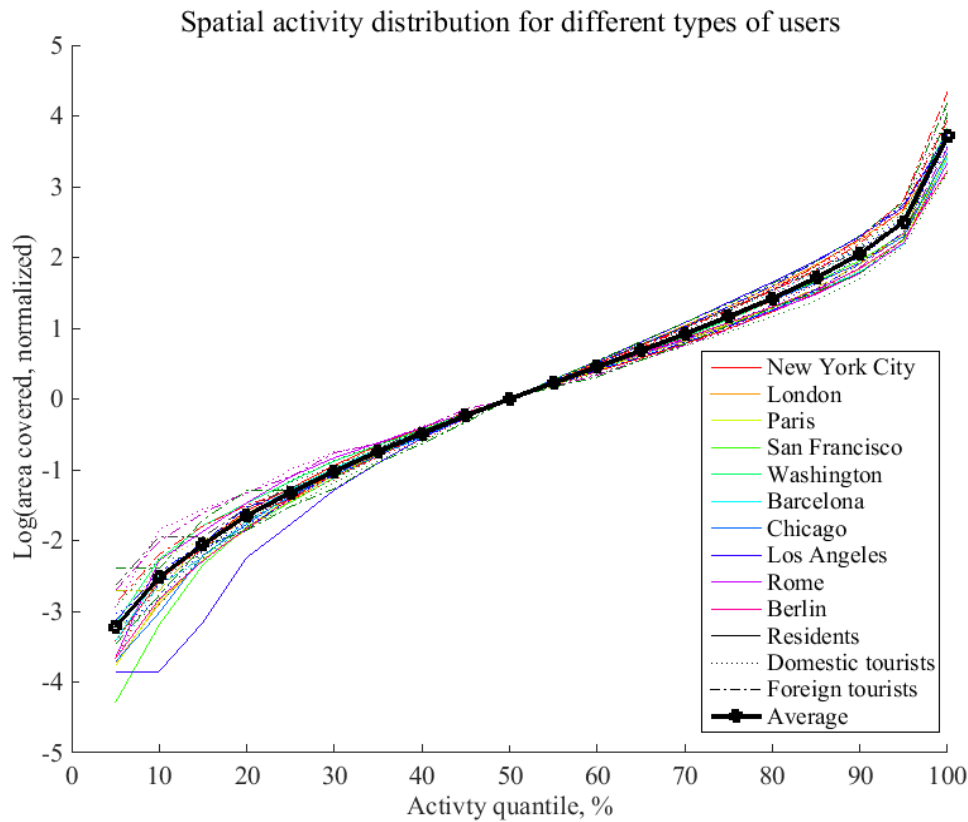

**Figure 7** – **Normalized spatial distribution.** Normalized spatial distribution of city attractiveness is shown for city residents, domestic tourists and foreign tourists – normalized total area of most photographed locations is covering a given quintile of the entire photographic activity (area covered by 50% of activity counted as a unit after normalization).

However, quantitatively the areas covered by the activity of city residents, domestic and foreign visitors within each city are quite different. Considering an average ratio between the size of areas covered by different activity quintiles (given the similarity of the spatial distribution shapes) for domestic/foreign tourists vs. the corresponding areas covered by the activity of residents for each city, we can see that these values vary largely from one city to another (see Table 4). The pattern is always the same – foreign tourist activity (in any given quintile) covers smaller areas compared to that one of domestic (the only exception is Berlin where those are almost the same) and, with the only exception of Los Angeles, the areas covered by both types of visitors are always smaller compared to the areas covered by the activity of residents. Once more, a very clear difference between the US and EU cities can be noticed – tourists visiting the US cities (with the exception of Chicago) explore more extensively the area compared to the ones visiting EU cities.

Table 4 – **Top photographed area.** Relative size of the top photographed area is covered by the activity of domestic and foreign visitors versus the corresponding area covered by residents in the different cities.

| City | Domestic tourists | Foreign tourists |
|---|---|---|
| **New York City** | 0.47 | 0.35 |
| **London** | 0.40 | 0.25 |
| **Paris** | 0.33 | 0.27 |
| **San Francisco** | 0.64 | 0.48 |
| **Washington** | 0.45 | 0.32 |
| **Barcelona** | 0.45 | 0.26 |
| **Chicago** | 0.37 | 0.21 |
| **Los Angeles** | 1.32 | 0.74 |
| **Rome** | 0.47 | 0.33 |
| **Berlin** | 0.25 | 0.26 |

The existing spatial activity coverage in each city is not homogenous as people are often initially attracted by its main destinations that become "must visit" locations consequentially getting the highest photographic activity. For city residents those locations, which are often called "hotspots", include parks, squares or sport facilities, while tourists are usually more attracted to the key places for the city identity (e.g., Times Square in New York City, Big Ben in London). The overall tourist activity in the New York City is visualized in Figure 8 by grouping coordinates of taken photographs into cells and creating a 3D map of the city perception. Our findings of New York City "hotspots" are similar to those previously identified in [22].

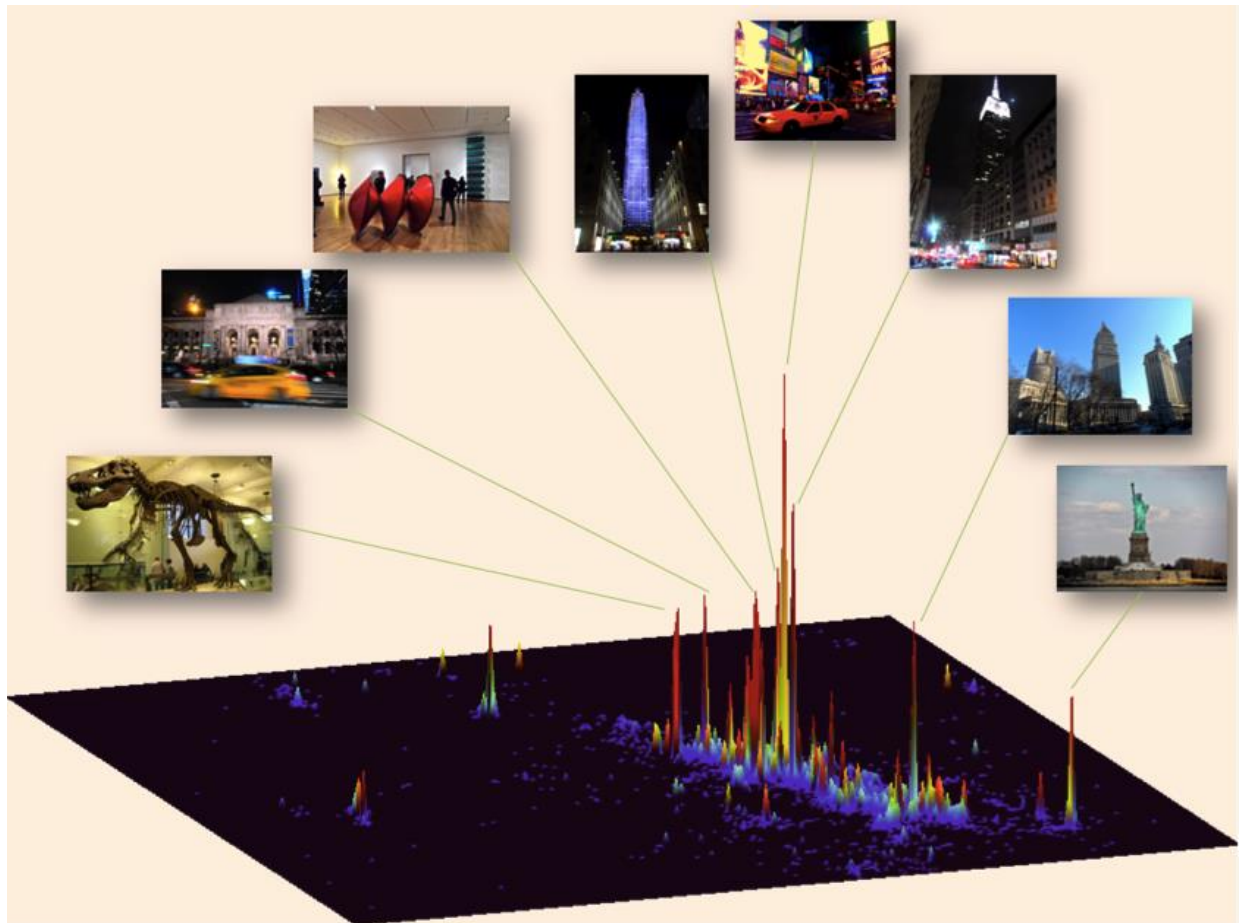

**Figure 8 – A tourist density attractiveness map of New York City.** The most photographed place in the city is Times Square. In Bronx, which does not attract many people, tourists are mostly taking photographs of Yankees stadium.

Further we give a definition of the hotspots as spatially connected areas on the rectangular grid which consist of the high-density cells and possess the highest cumulative activity. The given algorithm allows us to define a given number of spatial hotspots in all types of user activity, ordering them according to their cumulative activity level. Specifically we identify the top $n$ activity hotspots within each city using a following algorithm:

1. Consider top $n$ density cells in the grid and let $a$ be the lowest activity level among them.
2. Select all the cells with activity higher or equal to $a$ and divide them into spatially connected components (considering cells having at least one common vertex to be connected).
3. If resulting number of connected components is equal or higher than $n$ (could be higher if a number of cells has the same activity level $a$) stop the algorithm defining selected connected components as the hotspots.
4. Otherwise, if the number of components is $t < $ n, select n − t top activity cells from the remaining ones (not covered by selected components) and let $a$ be the lowest activity level among them. Repeat from Step 2.

Figure 9 reports the average total percentage of activity covered by n hotspots in our 10 cities depending on n. In the further analysis we will be considering $n = 12$ hotspots for each city on average covering around 30% of the total photographic activity of all users in the city, as a reasonable trade-off between having enough hotspots to represent important locations across the city and draw reliable conclusions from one hand, and the intent to have those hotspots cover just the major areas of interest across the city, but not majority of them from the other.

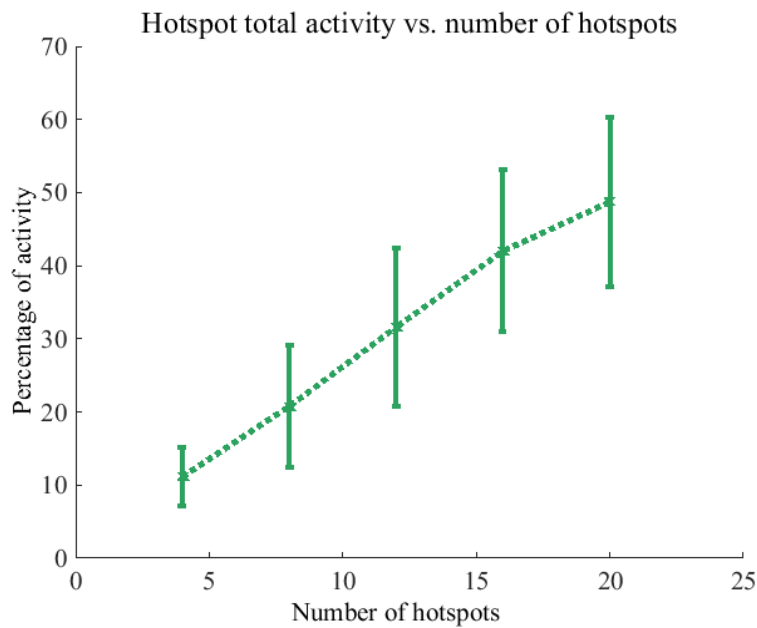

**Figure 9 − Hotspot activity coverage.** Average (together with the standard deviation) total hotspot activity per city depending on the number of hotspots.

In Figure 10 we show examples of the resulting maps for the total user activity (including residents and tourists) in one US city (i.e., New York City) and one EU city (i.e., Rome). In both cities the most active area is their downtown and moreover, one hotspot represents their stadiums – Yankee Stadium in New York City and Olimpico in Rome. Another similarity is that in both cities among their top 12 hotspots are their squares and parks, meaning that outside of their downtowns and besides their most famous attractions, people take photographs and spend their time at locations where they can meet and talk with other people. The only two cultural hotspots outside the downtown of Rome are the Church San Paolo Fuori le Mura and the Capitolium. Similarly, in New York City hotspots that are outside downtown are two museums located close to Central Park – the Metropolitan Museum of Art and the American Museum of Natural History. Table 5 gives the complete list of top 12 hotspots in Rome and New York City.

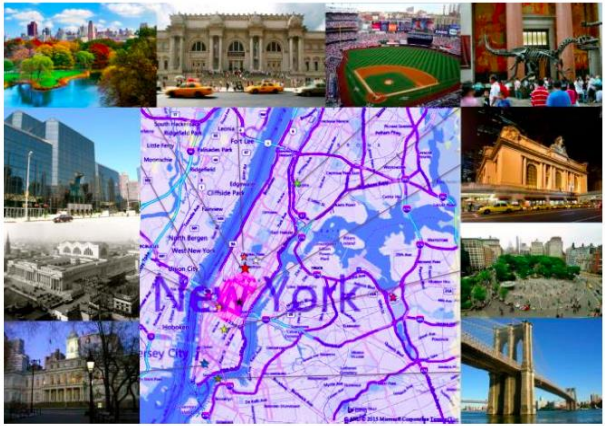 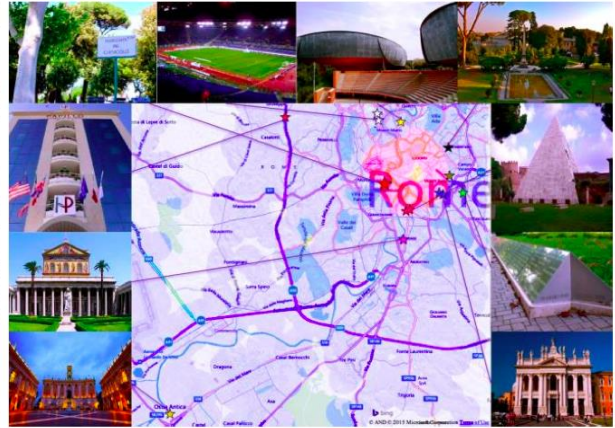

**Figure 10 –** **City maps of (a) New York City and (b) Rome.**

**Table 5 –** **Hotspots in New York City and Rome.**

| Hotspot rank | New York City | Rome |
|---|---|---|
| 1 | Downtown | Downtown |
| 2 | New York City Hall | San Giovanni in Laterano square |
| 3 | Metropolitan Museum of Art | Olimpico Stadium |
| 4 | Union Square | Auditorium Parco della Musica |
| 5 | Citi Field | Caio Cestio Pyramide |
| 6 | Yankee Stadium | IED – Istituto Europeo di Design |
| 7 | American Museum of Natural History | Papillo Hotel |
| 8 | Penn Station | Capitolium |
| 9 | Grand Central Terminal | Villa Torlonia |
| 10 | Javitz Center | Church San Paolo Fuori Le Mura |
| 11 | Brooklyn Bridge | Parco dei Caduti |
| 12 | Central Park | Passeggiata del Gianicolo |

Once that we identified the first dozen of the most photographed spatial hotspots for each city for all types of users, we define the cumulative activity within each of them separately for residents, domestic and foreign visitors and rank them in the decreasing order by the number of photographs taken. We then plot that number vs. the hotspot rank (similar to the approach used in González et al., 2008). Approximating these rank plots with power law dependences $hotspotActivity(rank) = c\, rank^{-q}$ (by doing so we get $R^2$ above 90% on average) gives us another important quantitative relative characteristic of how concentrated or distributed between the key destinations across the city is the activity of different types of people. Figure 11 shows those distributions on an example of New York City where the distribution is more flat (i.e., lower $q$) for the residents and sharper for domestic and especially foreign tourists (i.e., higher $q$).

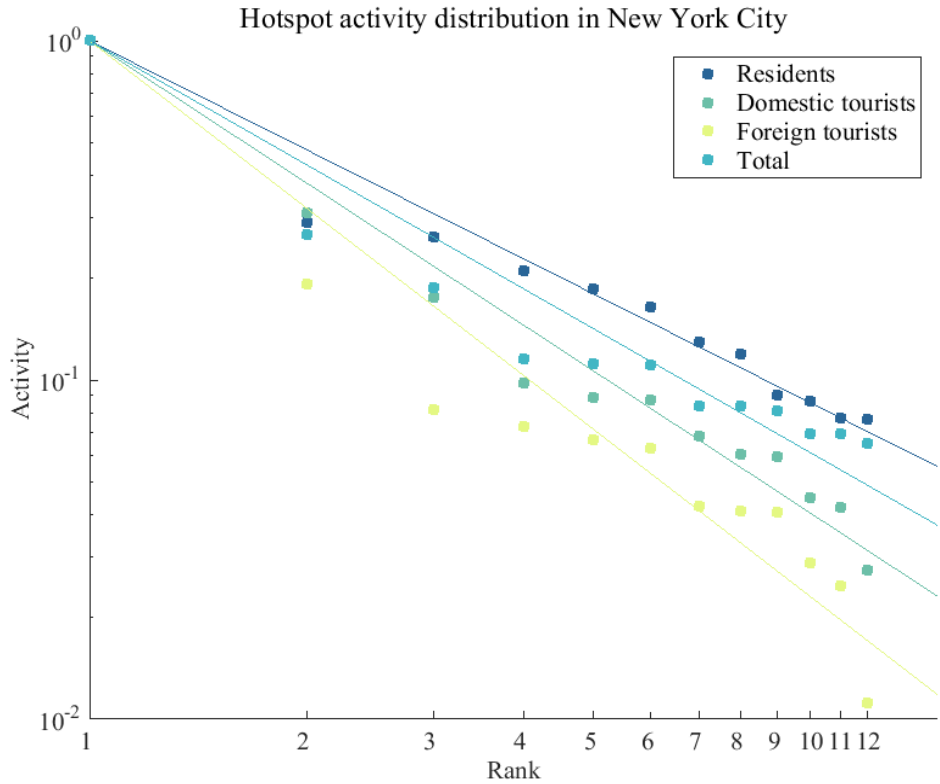

**Figure 11 – Log-log rank plot of 12 hotspots relative attractiveness in New York City fitted to power-laws.** The top hotspot attractiveness is considered as a unit for each type of users.

Detailed comparison of $q$ exponents for different cities is presented in Table 6. From the values of q we can see how concentrated or distributed resident and tourist activity is within them. If the value of $q$ exponent is higher, then it means that people are more focused on just visiting the major attractions, while a lower exponent means that people attention is more equally distributed among different attractions. By quantifying $q$ separately for residents, domestic and foreign tourists, we can conclude that their values largely vary from one city to another, corresponding to the unique city spatial layout. However, we can still discover additional strongly consistent pattern in human behavior across different cities: the distribution always becomes sharper for the visitors and especially for the foreign visitors with the only exception of Berlin where foreign activity has almost the same exponent as the activity of the city residents.

**Table 6 – Values of $q$ exponents for residents and tourists in the different cities.** $q$ exponent is a measure of how concentrated or distributed people activity is across different "hotspots". Namely, a higher value means that people are more focused on just visiting major attractions.

| *City* | *Residents* | *Domestic tourists* | *Foreign tourists* | *Total activity* |
|---|---|---|---|---|
| **New York City** | 1.06 | 1.39 | 1.63 | 1.21 |
| **London** | 1.29 | 1.35 | 1.61 | 1.25 |
| **Paris** | 1.25 | 1.38 | 1.43 | 1.23 |
| **San Francisco** | 1.10 | 1.12 | 1.21 | 1.02 |
| **Washington** | 1.21 | 2.51 | 2.81 | 1.42 |
| **Barcelona** | 1.65 | 2.18 | 2.26 | 1.69 |
| **Chicago** | 1.48 | 1.91 | 2.71 | 1.54 |
| **Los Angeles** | 1.25 | 1.46 | 1.84 | 0.92 |
| **Rome** | 2.37 | 2.77 | 3.17 | 2.61 |
| **Berlin** | 1.60 | 2.19 | 1.60 | 1.56 |

# 5 Summary and Conclusion

The importance of cities in our society is well founded and it is evident that cities play a crucial role as more than half of world population lives in them. "Rethinking" cities is thus the key component of the world sustainable development paradigm. The first and the most direct way of doing that is by rethinking the way we plan them. In order to become a better planner, one needs to start considering people needs. Not only do people need efficiency, better transportation and green energy, but also do they need a better experience of living in cities enjoying the things that they have interest in and that they find attractive. In this study we thus conducted an analysis of cities through the city attractiveness and derived patterns. The novelty of the study is in the kind of data that was used in it: geotagged photographs from publicly available photograph sharing web sites (e.g., Flickr).

Over the last decade big data analyses have being increasingly utilized in urban planning (e.g., analysis of cell phone records for the transportation planning). However, information from geotagged photographs was not very often analyzed although they can provide us with an additional layer of information useful for the urbanism in general. Namely, taken photographs indicate places in cities important enough for people to visit them and to decide to leave their digital trails there. By analyzing the global dataset of geotagged photographs we identified 10 most photographed cities, which happened to be distributed evenly between the US and EU. Focusing on the top 10 selected destinations, we studied spatial patterns of visitor attraction versus behavior of the residential users together with analyzing people mobility between those 10 cities and other places all around the world.

Although intercity origin/destination fluxes in a rather predictable way depend on the distance between two cities, links between American and European cities are surprisingly asymmetric. Namely, links going from American origins to EU destinations are on average stronger than the ones going in the opposite direction. Another clearly distinctive pattern between EU and American cities is related to the structure of the photographic activity within them. The results showed that in the US cities their residents take most of the photographs while the domestic tourists mostly cover the rest leaving not much for foreigners. The activity that happens in the EU cities is much more diverse showing a higher fraction of touristic and specifically foreign activity. Finally, when investigating the qualitative structure of destinations, again American and EU patterns are surprisingly distinctive – while Americans seem mostly engaged in domestic tourism, the Europeans travel more abroad than within their own home countries.

Moreover, we extracted the photographic activity at the local scale comparing attraction patterns for residents, domestic and foreign tourists within a city. Spatial distribution of photographic activity of all those user categories follows the same universal pattern – activity density distributions of all types of users follow log-normal law pretty well while the shapes of the curves for area size vs. activity quintile appear to be strongly consistent. However, the areas covered by tourist activities are always smaller compared to the areas covered by residents with the only exception of Los Angeles where domestic tourist activities cover larger areas compared to city residents. The ratio between areas covered by tourists and city residents is different for domestic and foreign tourists and is always higher for domestic tourists with only exception of Berlin where those factors are almost the same.

Once again, we find the activities within American and European cities are different from the quantitative standpoint. First, tourists visiting American cities (with the exception of Chicago) explore them more extensively, covering more of the areas of residential activity. More strikingly, the variance of the foreign activity density distribution is always higher for all the European cities compared to the residential activity and lower for all the American cities. Finally, we identified the hotspots in each city focusing on the most photographed places of each city. We noted that hotspot attractiveness follows a power-law distribution where the exponent of this distribution serves as an indicator of how focused people's attention is on the major attractiveness compared to how distributed it is among a number of objectives. For all cities with the exception of Berlin, activity of the tourists and especially foreign tourists appeared to be more concentrated on the major attractions.

To conclude, by showing differences between people visiting the US and EU cities our study revealed interesting patterns in human activity. The results of our study are useful for understanding of what has to be enhanced in cities and where it can be appropriate to increase services targeting different categories of users. In past those questions were traditionally answered by analyzing different available datasets such as hotel information or survey data. However, collecting or getting the access to such datasets usually requires significant efforts and/or expenses, while geotagged photography is publicly available while providing a unique global perspective on addressing many research questions at both global and local scale. In future work we will also consider the longitudinal perspective of data analysis by showing how the observed human patterns evolve over time.

**Competing interests**

The authors declare that they have no competing interests.

**Authors' contributions**

MG, SS and CR defined the research agenda, SP and IB processed the data, SP, IB and SS analyzed the results, SP, IB and SS wrote the manuscript, all authors edited the manuscript.

**Acknowledgements**

The authors wish to thank MIT SMART Program, Accenture, Air Liquide, BBVA, The Coca Cola Company, Emirates Integrated Telecommunications Company, The ENEL foundation, Ericsson, Expo 2015, Ferrovial, Liberty Mutual, The Regional Municipality of Wood Buffalo, Volkswagen Electronics Research Lab, UBER and all the members of the MIT Senseable City Lab Consortium for supporting the research. We further thank MIT research support committee via the NEC and Bushbaum research funds, as well as the research project "Managing Trust and Coordinating Interactions in Smart Networks of People, Machines and Organizations", funded by the Croatian Science Foundation. Finally, we would like to thank Eric Fisher for providing the dataset for this research, Alexander Belyi for some help with the visualizations and also Jameson L Toole, Yingxiang Yang, Lauren Alexander for useful recommendations at the early stage of this work.